\documentclass[conference,9pt]{IEEEtran}

\AtBeginDocument{%
  \providecommand\BibTeX{{%
    \normalfont B\kern-0.5em{\scshape i\kern-0.25em b}\kern-0.8em\TeX}}}

\usepackage{titlesec}
\usepackage{enumitem}
\usepackage{microtype}
\usepackage{graphicx}
\usepackage{booktabs}
\usepackage{tikz}
\usepackage{url}
\usepackage{amsmath}
\usepackage{mathtools}
\usepackage{subcaption}
\usepackage{amsthm}
\usepackage{textcomp}
\usepackage{amsmath}
\DeclareMathOperator*{\argmax}{arg\,max}

\usepackage[capitalize,noabbrev]{cleveref}

\theoremstyle{plain}

\theoremstyle{definition}

\theoremstyle{remark}

\usepackage[textsize=tiny]{todonotes}
\usepackage{multirow}
\usepackage{graphicx}
\usepackage{caption}
\usepackage{longtable}

\begin{document}

\newcommand{\pgftextcircled}[1]{
    \setbox0=\hbox{#1}%
    \dimen0\wd0%
    \divide\dimen0 by 2%
    \begin{tikzpicture}[baseline=(a.base)]%
        \useasboundingbox (-\the\dimen0,0pt) rectangle (\the\dimen0,1pt);
        \node[circle,draw,outer sep=0pt,inner sep=0.1ex] (a) {#1};
    \end{tikzpicture}
}


\title{Mapping Fusion: Improving FPGA Technology Mapping with ASIC Mapper}


\author{
    Cunxi Yu \\
    University of Maryland, College Park \\
    Maryland, USA \\
    cunxiyu@umd.edu
}

\maketitle
\begin{tikzpicture}[remember picture,overlay]
\node at (current page.south) [yshift=10pt] {
    \parbox{\textwidth}{
        \centering
        \footnotesize
        979-8-3315-3762-3/25/\$31.00~\copyright~2025 IEEE
    }
};
\end{tikzpicture}

\begin{abstract}

LUT (Look-Up Table) mapping is a critical step in FPGA logic synthesis, where a logic network is transformed into a form that can be directly implemented using the FPGA's LUTs. 
The goal of LUT mapping is to map the Boolean network into LUTs, where each LUT can implement any function with a fixed number of inputs. 
In parallel to FPGA technology mapping, ASIC technology mapping maps the Boolean network to user-defined standard cells, which has traditionally been developed separately from LUT mapping algorithms. However, in this work, our motivating examples demonstrate that ASIC technology mappers can potentially improve the performance of LUT mappers, such that standard cell mapping and LUT mapping work in an incremental manner. 

Therefore, we propose the FuseMap framework, which explores this opportunity to improve LUT mapping in the FPGA design flow by utilizing reinforcement learning to make design-specific choices during cell selection. 
The effectiveness of FuseMap is evaluated on a wide range of benchmarks, different technology libraries, and technology mappers. The experimental results demonstrate that FuseMap achieves higher mapping accuracy while reducing delay and area across diverse circuit designs collected from ISCAS 85/89, ITC/ISCAS 99, VTR 8.0, and EPFL benchmarks. 

\end{abstract}


\section{Introduction}

Field-Programmable Gate Arrays (FPGAs) have emerged as versatile, high-performance platforms for a wide range of applications, from data centers to edge computing. Their inherent reconfigurability allows hardware designers to rapidly prototype and deploy systems, making them highly attractive for applications where flexibility and customizability are critical. Unlike Application-Specific Integrated Circuits (ASICs), which are fixed-function, FPGAs offer the ability to be reprogrammed, allowing users to map custom designs onto general-purpose hardware.

At the heart of FPGA design lies the concept of \textbf{Look-Up Tables (LUTs)}, which are fundamental building blocks used to implement logic functions. Each LUT can be configured to represent any Boolean function of a given number of inputs $K$, making them highly flexible in expressing combinational logic. However, efficient LUT mapping, where the design is converted into an optimized network of LUTs, remains a challenging task. The quality of this mapping process directly influences the overall performance, area, and power consumption of the final FPGA implementation. Over the past few decades, LUT mapping for FPGAs has been a significant area of research aimed at improving performance, area efficiency, and power consumption. The field has explored LUT mapping from basic area/delay optimization to advanced techniques incorporating physical and power awareness, as well as emerging technology-based FPGAs \cite{cong1993area,cong1994flowmap,farrahi1994complexity,chen2004low,mishchenko2006improvements,ling2005fpga,cong1998technology}.

As FPGA technology and LUT mapping algorithms have evolved over the years, ASIC technology mapping has also seen significant advancements. However, there is no prior work exploring the potential integration of ASIC standard-cell technology mapping flows with LUT mapping. Intuitively, packing logic into standard cells could provide a better balance between functionality and structure in LUT minimization. Moreover, as LUT sizes increase to $K=6$, which is larger than most modern standard cells, ASIC mapping could serve as a pre-packing phase to improve LUT mapping performance. The optimization potential is first confirmed by our motivation case studies in Section \ref{sec:case_study}, and is followed by the development of our proposed FuseMap framework.

The main contributions of this work are summarized as follows: (1) The first case studies of fusing modern ASIC technology mapping with LUT mapping to improve FPGA LUT mapping results are presented in Section \ref{sec:case_study}. (2) Motivated by the case studies, we present a novel lightweight reinforcement learning-based approach to optimize the entire fused mapping process. The key challenge in this approach is fine-tuning the technology libraries in ASIC mapping to improve LUT mapping results. (3) FuseMap is evaluated using a variety of benchmarks from ISCAS 85/89/99 \cite{brglez1989combinational}, VTR \cite{luu2014vtr}, and EPFL benchmarks \cite{soeken2018epfl}, which demonstrates consistent improvements in LUT minimization. The source code of this work can be found at \texttt{``lut``} branch at \url{https://github.com/Yu-Maryland/MapTune}.
\section{Background}

\subsection{Boolean Networks and AIGs}
\label{sec:background_BN}
A Boolean network is a directed acyclic graph (DAG) denoted as $G=(V, E)$ with nodes $V$ representing logic gates (Boolean functions) and edges $E$ representing the wire connection between gates. The input of a node is called its \textit{fanin}, and the output of the node is called its \textit{fanout}. The node $v \in V$ without incoming edges, i.e., no \textit{fanins}, is the \textit{primary input} (PI) to the graph, and the nodes without outgoing edges, i.e., no \textit{fanouts}, are \textit{primary outputs} (POs) to the graph. The nodes with incoming edges implement Boolean functions. The level of a node $v$ is defined by the number of nodes on the longest structural path from any PI to the node inclusively, and the level of a node $v$ is noted as $level(v)$.

\textit{And-Inverter Graph} (AIG) is one of the typical types of DAGs used for logic manipulations, where the nodes in AIGs are all two-inputs AND gates, and the edges represent whether the inverters are implemented. An arbitrary Boolean network can be transformed into an AIG by factoring the sum-of-products (SOPs) of the nodes, and the AND gates and OR gates in SOPs are converted to two-inputs AND gates and inverters with DeMorgan's rule. There are two primary metrics for evaluation of an AIG, i.e., \textit{size}, which is the number of nodes (AND gates) in the graph, and \textit{depth}, which is the number of nodes on the longest path from PI to PO (the largest level) in the graph. 
{A \textit{cut} $C$ of node $v$ includes a set of nodes of the network. The leaf nodes included in the \textit{cut} of node $v$ are called \textit{leaves}, such that each path from a PI to node $v$ passes through at least one leaf. The node $v$ is called the \textit{root} node of the \textit{cut} C. The cut size is the number of its leaves. A cut is $K$-feasible if the number of leaves in the cut does not exceed $K$.}

\subsection{FPGA LUT Mapping}


LUT mapping is the process of translating a given Boolean network into a network composed of $k$-input LUTs ($k$-LUTs), where $k$ is the maximum number of inputs each LUT can handle. This transformation is a crucial step in FPGA design, as it impacts both performance and resource utilization. LUT mapping begins with a Boolean network represented in a form that allows easy manipulation, such as an And-Inverter Graph (AIG). 
The goal of LUT mapping is to identify subsets of the graph that can be collapsed into single LUTs, ensuring that the entire network is covered by a minimal set of LUTs.

The core of the LUT mapping process involves three main steps: \textit{Cut enumeration}: For each node in the AIG, cuts are computed, representing subsets of the graph that can potentially be mapped to a LUT. \textit{Mapping}: From the set of computed cuts, a selection is made that ensures full coverage of the Boolean network while minimizing a specific cost function. This function may target metrics such as delay, area, or power consumption. \textit{Post-optimization}: After the initial mapping, optimizations can also be applied to the LUT network to improve the quality of results (QoR), aiming at further reducing the number of LUTs, improve the critical path delay, or optimize for other design goals.


\subsection{ASIC Technology Mapping and Library}

Standard-cell technology mapping for ASIC design is another key topic in logic synthesis process, converting high-level circuit descriptions into technology-specific gate-level netlists, particularly for Application-Specific Integrated Circuits (ASIC) designs. It involves selecting appropriate cells from an EDA library to realize a circuit in a chosen technology, effectively bridging high-level design with physical implementation. 

Various algorithms have been developed to address the technology mapping problem, crucial in logic synthesis for ASIC design. These include tree-based approaches \cite{marwedel1993tree,cong1994flowmap}, which focus on mapping trees or sub-trees to specific gates, and Directed Acyclic Graph (DAG)-based methods \cite{mishchenko2006improvements,mishchenko2005technology,mishchenko:2006-dag} that consider the entire circuit topology for enhanced optimization. 
Recently, ML approaches have also been involved in improving the technology mapping process, either by correlating the technology-independent representation to technology-dependent PPAs (prediction models) \cite{neto2021read,li2024boolgebra}, or by directly optimizing the technology mapping algorithms \cite{neto2021slap}. 


Technology libraries are critical for ASIC mapping. Consequently, significant efforts have been made to optimize or generate standard cell libraries to improve the power, performance, and area (PPA) metrics in the design flow \cite{ren2021nvcell}. However, there has been little to no effort to analyze the impact of these libraries on the technology mapping procedure itself. Specifically, in this work, we observe and demonstrate an interesting finding: \textit{a partially selected set of cells from the full technology library can improve the LUT mapping results in fused mapping setups, whereas using the full library for ASIC mapping tends to worsen the LUT mapping results}. A more comprehensive case study in Section \ref{sec:case_study} will discuss the fused mapping performance under default ASIC technology mapping setups versus partially selected library mapping.

\subsection{Machine Learning for EDA}

Recent years have seen increased use of decision intelligence in EDA to reduce manual effort and accelerate design closure in modern toolflows \cite{kapre2015intime,ustun2019lamda,ziegler2017ibm,DBLP:conf/dac/YuXM18,hosny2019drills,ma2019high,yu2020decision,yu2020flowtune}. Machine learning (ML) techniques have been applied to automatically configure industrial FPGA \cite{kapre2015intime,ustun2019lamda,neto2022flowtune,pal2022machine}, ASIC toolflows \cite{ziegler2017ibm, li2016efficient, DBLP:conf/dac/YuXM18, hosny2019drills, liu2024maptune, li2023verilog,chen2024syn}, formal reasoning \cite{wu2023gamora,deng2024less,cai2025smoothe}, and compilation optimization \cite{liu2024differentiable,yin2023respect,yin2023accelerating,cai2025smoothe}. Synthesis flow exploration has also been enhanced using CNNs \cite{DBLP:conf/dac/YuXM18} and reinforcement learning \cite{hosny2019drills}, improving quality-of-results \cite{DBLP:conf/dac/YuXM18,hosny2019drills}. Additionally, lightweight MAB models effectively identify optimal synthesis flows, balancing exploration and exploitation \cite{neto2022flowtune,liu2024cbtune}. 

\section{Motivating Case Studies -- Fused Mapping}\label{sec:case_study}

 
This section presents the first exploration of fusing ASIC and LUT mapping to motivate this work. The case study results are presented in Figure \ref{fig:casestudy}. All results of the conducted case studies are based on synthesis framework ABC \cite{mishchenko2007abc} with various ASIC mapping options within \texttt{map} and LUT mapper \texttt{if -K} followed by post-optimization \texttt{mfs2}. 
By undertaking this case study, we aimed to investigate optimization opportunities through fused mapping process. Our analysis primarily focused on design-specific scenarios, aiming at providing valuable insights into the selection of an appropriate ASIC technology mapping configurations for improving LUT mapping results. Based on the experimental results summarized in Figure \ref{fig:casestudy}, three critical observations have been summarized:

\begin{figure}
    \centering
\includegraphics[width=0.5\textwidth]{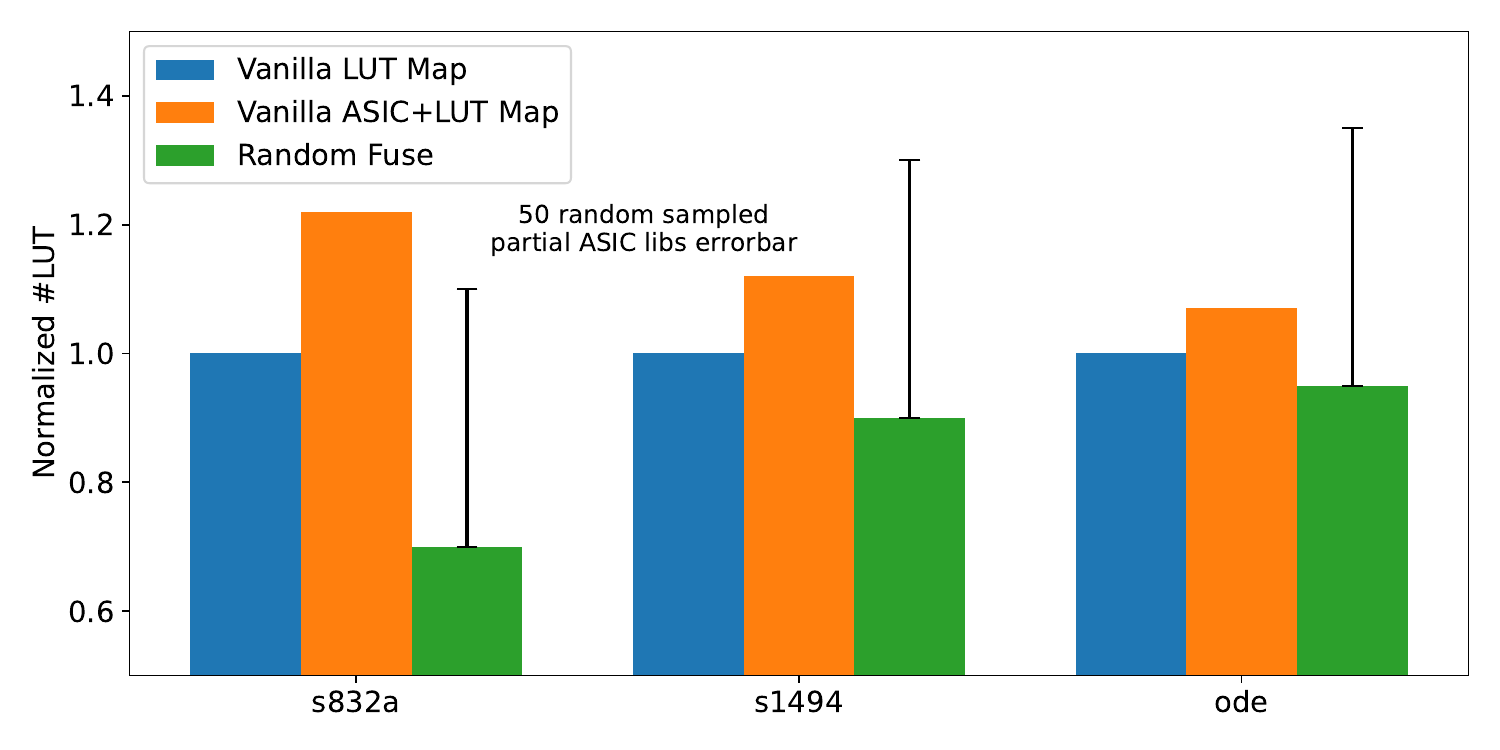}
            \caption{Motivation case studies of fusing ASIC and LUT mapping for $K=6$ using \texttt{ASAP7} library, conducted on delay-driven and area-driven mapping in ABC \texttt{map} and ABC LUT mapper \texttt{if -K}. The randomly partial sampled library contains 80 out of the original 161 ASAP7 cells, and this case study include 50 random samples.}
    \vspace{-2mm}
    \label{fig:casestudy}
\end{figure}

\noindent
\textbf{Observation 1 -- Simply fusing a vanilla LUT mapping with a vanilla ASIC mapping does not offer any improvements}. Compared to the baseline (vanilla LUT mapping) using ABC, simply combining an ASIC mapper with a LUT mapper as a flow to produce LUT netlists results in no improvements. As shown in Figure \ref{fig:casestudy}, the number of LUTs (\#LUT, representing area) consistently increases by 7\% to 24\%. 

\textbf{Observation 2 -- Optimization opportunities exist in fused mapping, allowing designs to be improved in terms of \#LUT by selectively including ASIC library cells.}
To further explore the potential of fused mapping, we examined whether the selection of technology standard cells in the fused mapping process makes a difference. Specifically, we randomly selected 80 cells out of the 161 cells available in the ASAP7 7nm library and performed the same fused mapping as the vanilla fused mapping (Vanilla ASIC+LUT). As shown in Figure \ref{fig:casestudy}, a key observation is that fused mapping not only improves upon straightforward vanilla fused mapping but also surpasses Vanilla LUT mapping, which represents the state-of-the-art (SOTA) open-source LUT mapper \cite{brayton2010abc}. 

\noindent
While fused mapping can yield significant benefits—up to a 30\% reduction in \#LUT for design \texttt{s832a}—the majority of designs exhibit a deterioration in either \#LUT, delay, or both. For example, in design \texttt{s838a}, while the best result achieves a 30\% \#LUT reduction compared to the SOTA LUT mapper, the average results are worse. In fact, the worst result leads to a 10\% increase in \#LUT. Similarly, designs \texttt{ode} and \texttt{s1494} show an average increase in \#LUT across the 50 randomly selected libraries. These observations indicate that fused mapping requires careful selection of standard library cells in the flow, serving as the primary motivation for this work.

\noindent
\textbf{Take-away} -- From the three aforementioned observations, we can draw an initial conclusion that partially sampled libraries for technology mapping are likely to achieve better results under the fusion mapping flow. This suggests that narrowing down the library size for this specific design does not entirely benefit technology mapping. Additionally, we must consider another potential drawback of smaller sampling sizes: while smaller sizes offer greater potential for improved results, there is a risk of failed mapping due to the reduced number of available components, which limits the ability to find a suitable mapping solution for certain designs. Therefore, it is crucial to carefully evaluate the trade-off between optimization objectives and the probability of successful mapping when determining the appropriate sampling size.

Through this case study, we first demonstrated the potential of improving the LUT mapping results by fusing ASIC mapping into the flow, and also highlighted the impact of partially sampled libraries on the potential for achieving better results. Our observations underscore the importance of tuning the technology mapping library to meet design-specific requirements. 

\section{Approach}\label{section:approach}

The proposed FuseMap framework aims to enhance LUT mapping by integrating ASIC mapping strategies into the traditional LUT mapping flow. At the core of the framework is a multi-armed bandit (MAB)-based approach leveraged from our previous work MapTune \cite{liu2024maptune}, which uses adaptive probability updates to select portions of a technology library that can potentially improve mapping results. The process begins with a Vanilla LUT mapping, which serves as the baseline target for performance comparisons during the reward update process. Simultaneously, the probability distribution $P(i)$ is initialized randomly with respect to a given full standard-cell library, containing $N$ cells. The framework then proceeds with two mapping flows: the ASIC Mapping Flow, which uses the full standard-cell library, and the LUT Mapping Flow, which serves as a comparison point to assess performance improvements in terms of LUT size and delay. 

The core of the framework is the Fusion Map Tuning module, where two different MAB strategies running simultaneously, MAB-$\epsilon$ and MAB-UCB, adjust the probabilities associated with selecting cells from either the full standard-cell library or a partially selected library. Based on the outcomes (rewards) from the mapping flows, the probabilities $P(i)$ are updated iteratively to refine the selection of cells from the library. The goal is to maximize the rewards, specifically to improve both the LUT size and delay metrics, through iterative updates of the library. The process terminates after a user-defined number of iterations for updating $P$, ensuring that the exploration-exploitation process converges to an optimal mapping configuration. Ultimately, the FuseMap framework explores a balanced trade-off between area- and delay-driven mappings, optimizing the final LUT mapping results by selectively utilizing cells from a partial standard-cell library.

\begin{figure}
    \centering
    \includegraphics[width=1\columnwidth]{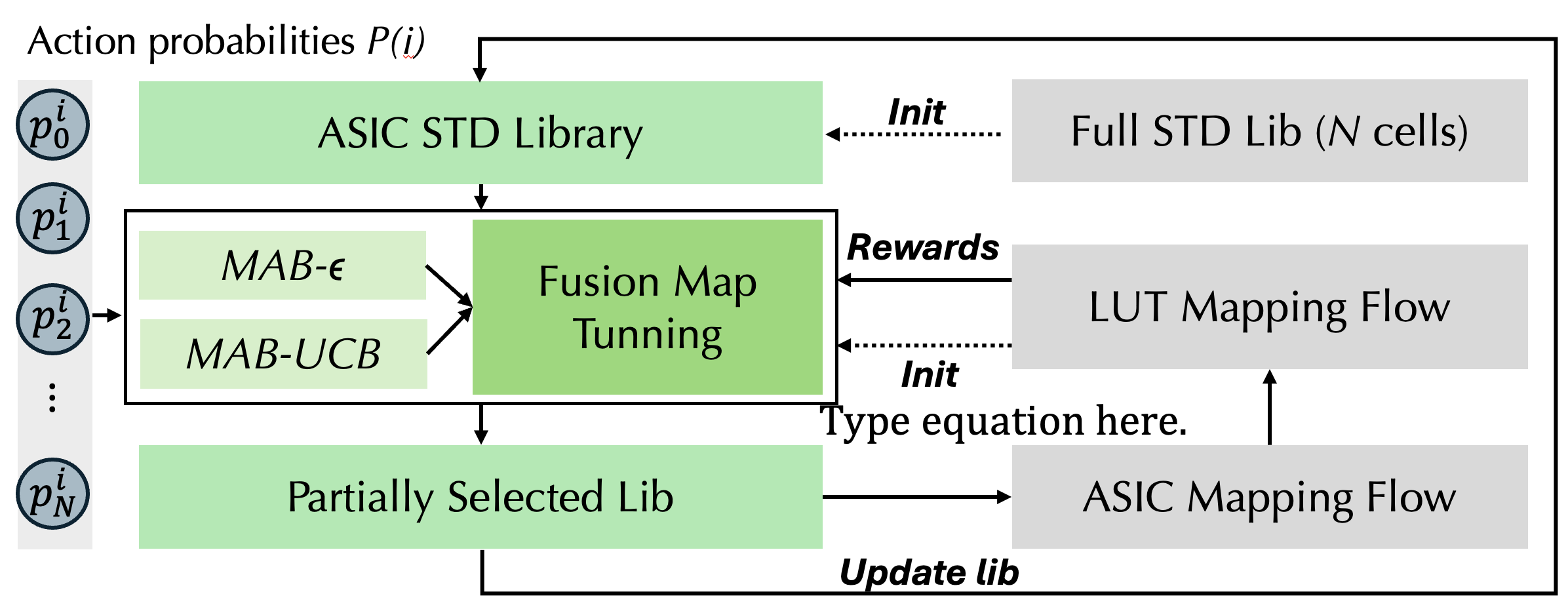}
    \caption{Overview of the FuseMap Framework. Termination is based on the user-defined number of iterations for updating $P$. The initialization process involves performing a Vanilla LUT mapping, which is set as the baseline target for reward updates. The probability $P$ is randomly initialized with respect to a given full standard-cell library.}
    \vspace{-3mm}
    \label{fig:flow}
\end{figure}
 
\begin{table}[h]
\centering
\caption{Notations of FuseMap formulation.}
\resizebox{\columnwidth}{!}{
\begin{tabular}{|c|l|}
\hline
\textbf{Notation} & \textbf{Description} \\
\hline
$\mathcal{L}$ & Set of all cell variants in the Library file \\
$\mathcal{N}$ & Number of cells variants in the Library file \\
$\mathcal{A}$ & Action space with a discrete and finite set of actions \\
$a^i$ & Action that select cell variant $i$\\
$\mathcal{S}$ & A state indicates which actions have been taken/which cell variants have been chosen \\
$\mathcal{R}_{\mathcal{S}}$ & Normalized reward under current state $\mathcal{S}$\\
$p_{a^i}$ & Probability of cell variant $i$ minimizing $\mathcal{R}$ \\

\hline
\end{tabular}}
\label{tab:notation}
\end{table}
 
\subsection{Formulation of FuseMap}


\begin{figure*}[ht]
    \centering
    
    \begin{subfigure}[t]{0.95\textwidth}
        \centering
        \includegraphics[width=\textwidth]{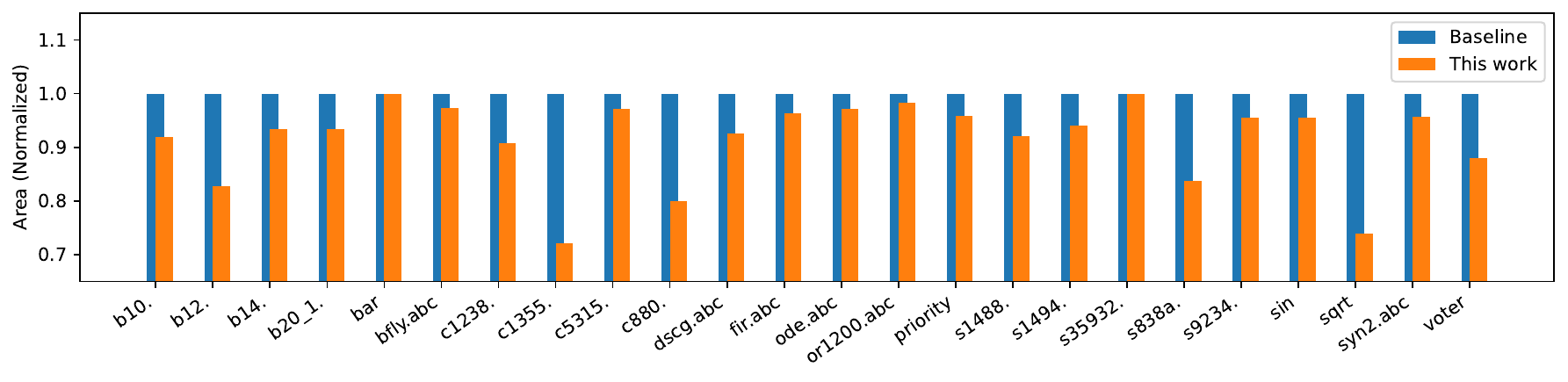}
        \caption{LUT size minimization as reward $\mathcal{R}$ in FuseMap using ASAP7 as initialized library.}
        
        \label{fig:7nm_area_comparison}
    \end{subfigure}
     
    \begin{subfigure}[t]{0.95\textwidth}
        \centering
        \includegraphics[width=\textwidth]{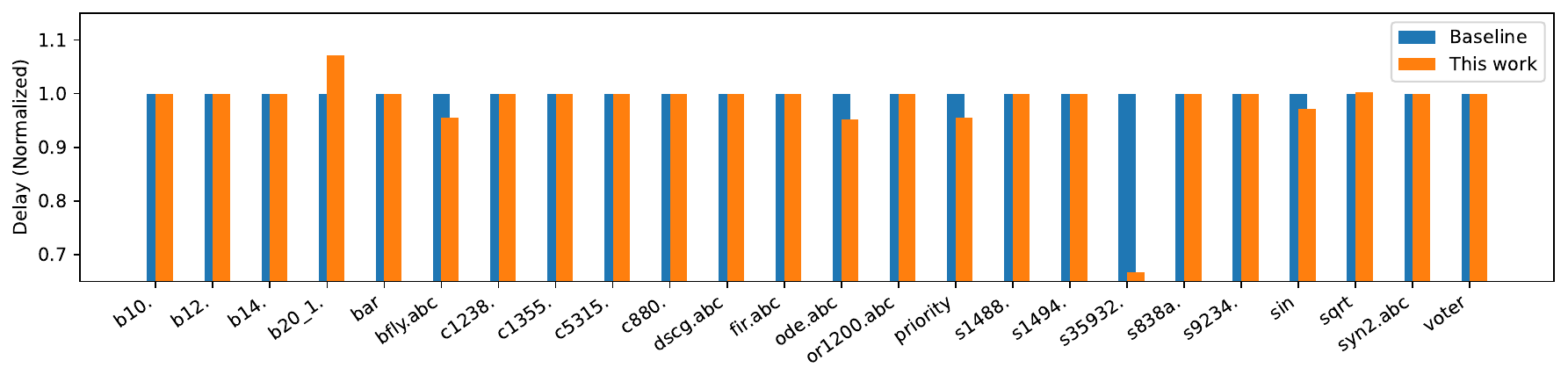}
        \caption{LUT depth minimization as reward $\mathcal{R}$ in FuseMap using ASAP7 as initialized library.}
        \vspace{-1mm}
        \label{fig:45nm_area_comparison}
    \end{subfigure}
 
    \begin{subfigure}[t]{0.95\textwidth}
        \centering
        \includegraphics[width=\textwidth]{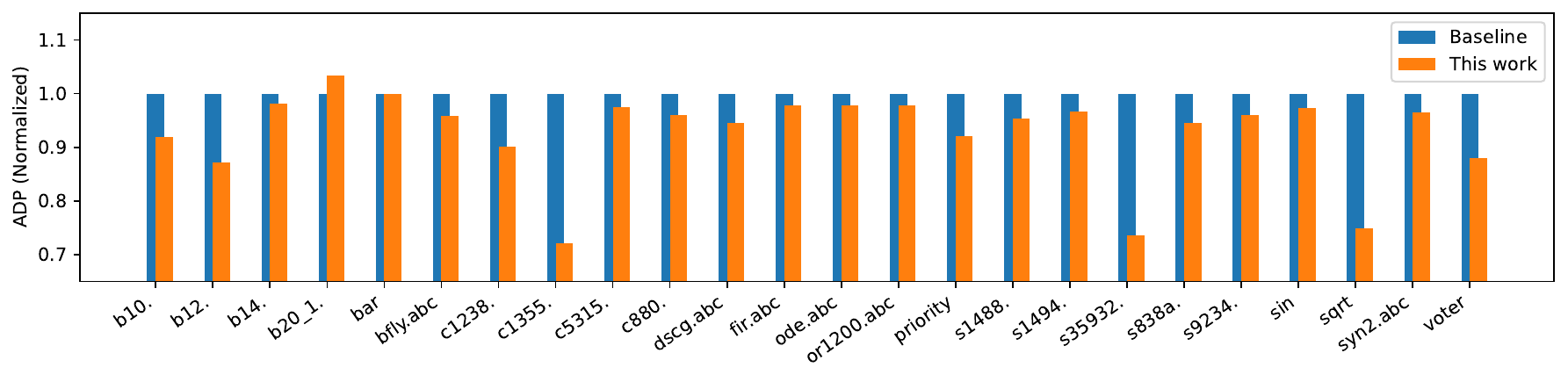}
        \caption{LUT network ADP (\#LUT size $\times$ depth) as reward $\mathcal{R}$ in FuseMap using ASAP7 as initialized library.}
         \label{fig:best_library_comparison}
    \end{subfigure}

    \caption{FuseMap results with three different reward functions over 24 designs using ASAP7 library.}
    \label{fig:fuse_map_all}
\end{figure*}

\begin{figure*}[ht]
    \centering
    
    \begin{subfigure}[t]{0.24\textwidth}
        \centering
        \includegraphics[width=\textwidth]{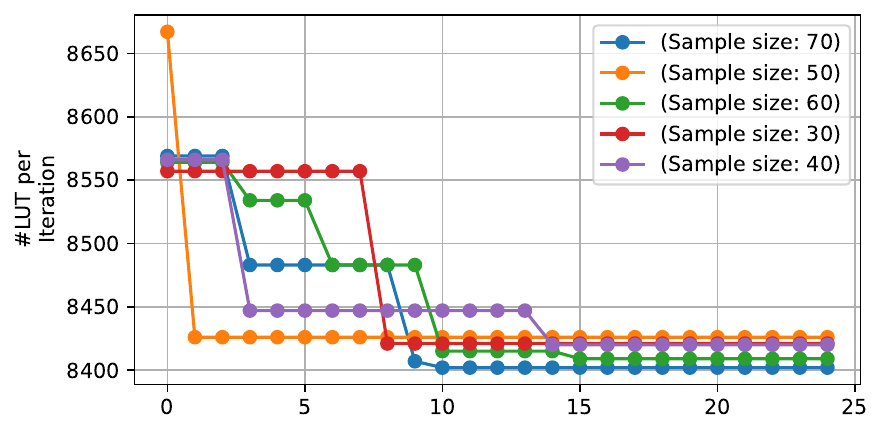}
        \caption{Design: bfly}
        \label{fig:b10}
    \end{subfigure}
    \hfill
    \begin{subfigure}[t]{0.24\textwidth}
        \centering
        \includegraphics[width=\textwidth]{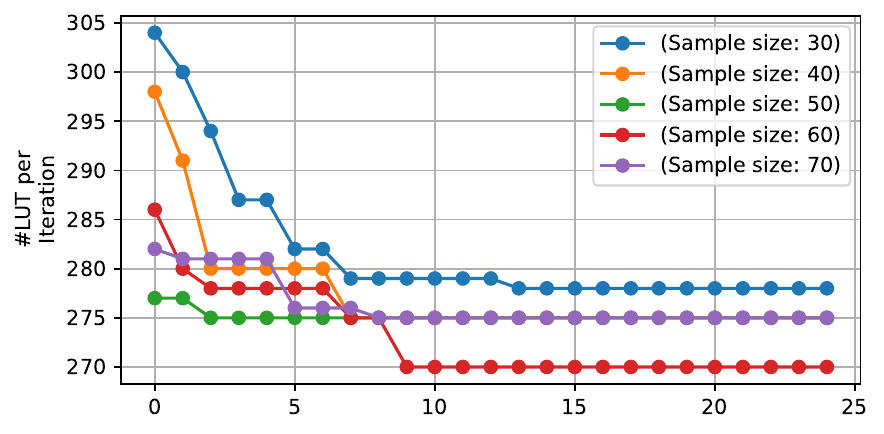}
        \caption{Design: b12}
        \label{fig:b12}
    \end{subfigure}
    \hfill
    \begin{subfigure}[t]{0.24\textwidth}
        \centering
        \includegraphics[width=\textwidth]{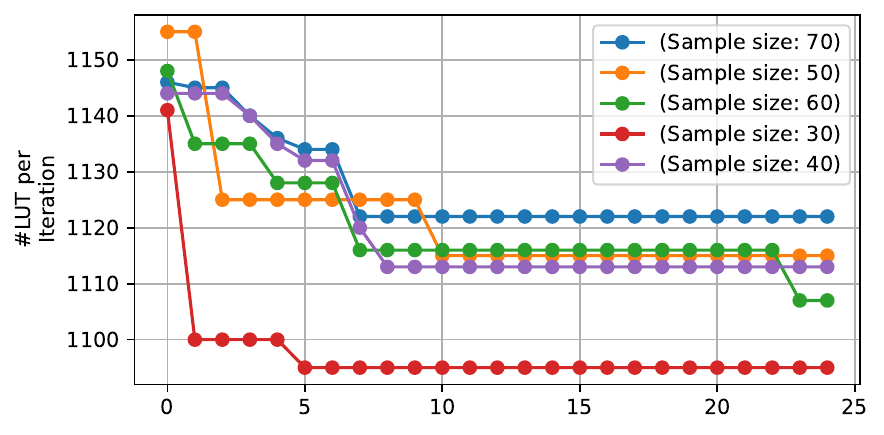}
        \caption{Design: b14}
        \label{fig:b14}
    \end{subfigure}
    \hfill
    \begin{subfigure}[t]{0.24\textwidth}
        \centering
        \includegraphics[width=\textwidth]{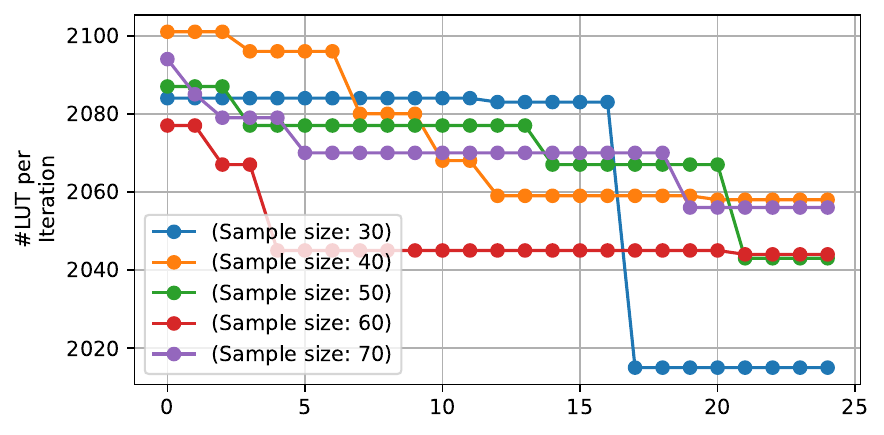}
        \caption{Design: b20\_1}
        \label{fig:b20_1}
    \end{subfigure}
    \begin{subfigure}[t]{0.24\textwidth}
        \centering
        \includegraphics[width=\textwidth]{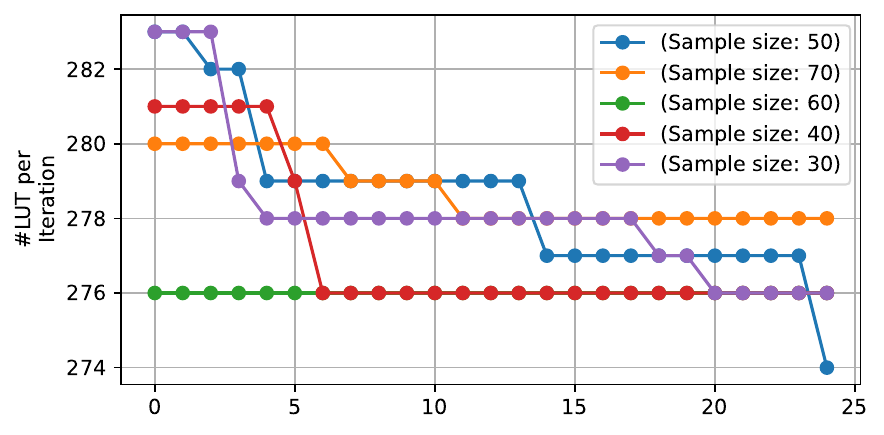}
        \caption{Design: c5315}
        \label{fig:c5315}
    \end{subfigure}
    \hfill
    \begin{subfigure}[t]{0.24\textwidth}
        \centering
        \includegraphics[width=\textwidth]{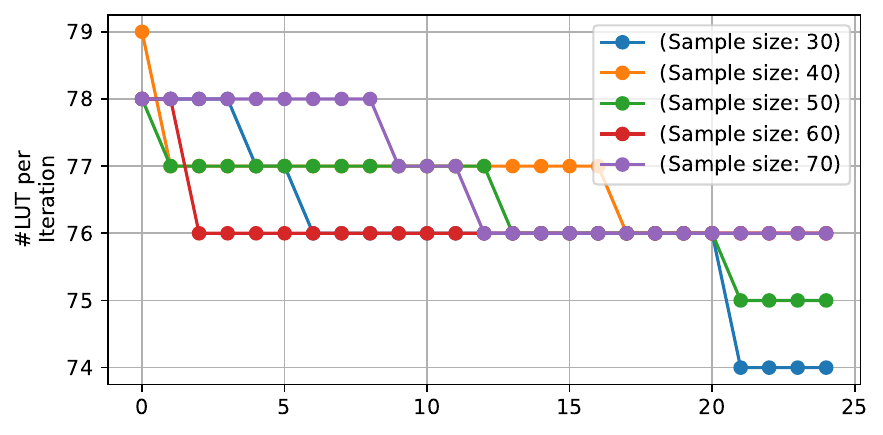}
        \caption{Design: c880}
        \label{fig:c880}
    \end{subfigure}
    \hfill
    \begin{subfigure}[t]{0.24\textwidth}
        \centering
        \includegraphics[width=\textwidth]{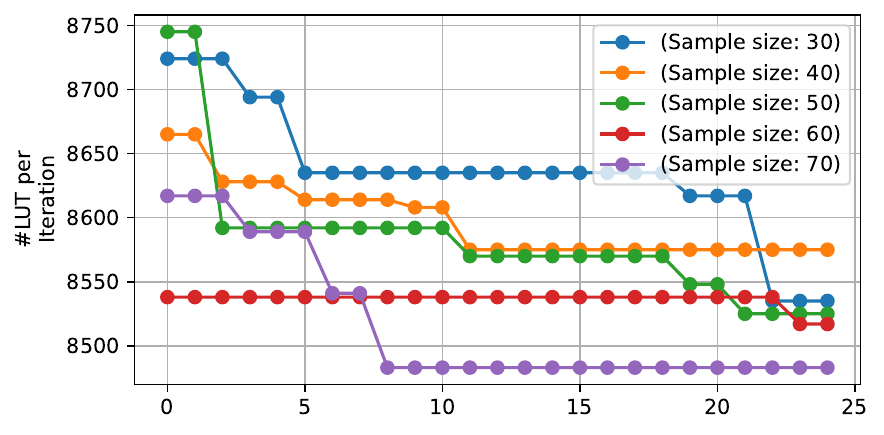}
        \caption{Design: dscg.abc}
        \label{fig:dscg_abc}
    \end{subfigure}
    \hfill
    \begin{subfigure}[t]{0.24\textwidth}
        \centering
        \includegraphics[width=\textwidth]{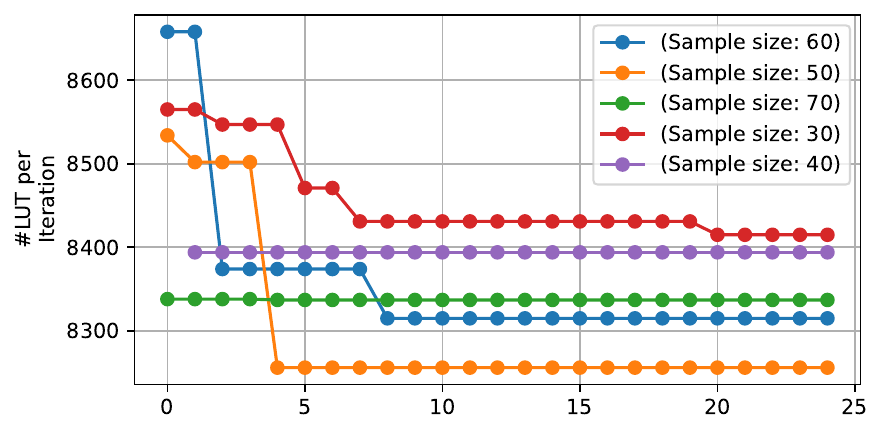}
        \caption{Design: fir.abc}
        \label{fig:fir_abc}
    \end{subfigure}
    
    \begin{subfigure}[t]{0.24\textwidth}
        \centering
        \includegraphics[width=\textwidth]{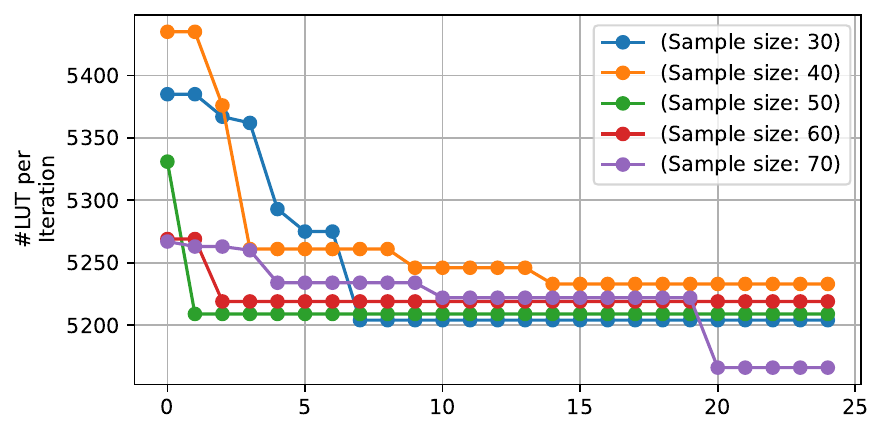}
        \caption{Design: ode.abc}
        \label{fig:ode_abc}
    \end{subfigure}
    \hfill
    \begin{subfigure}[t]{0.24\textwidth}
        \centering
        \includegraphics[width=\textwidth]{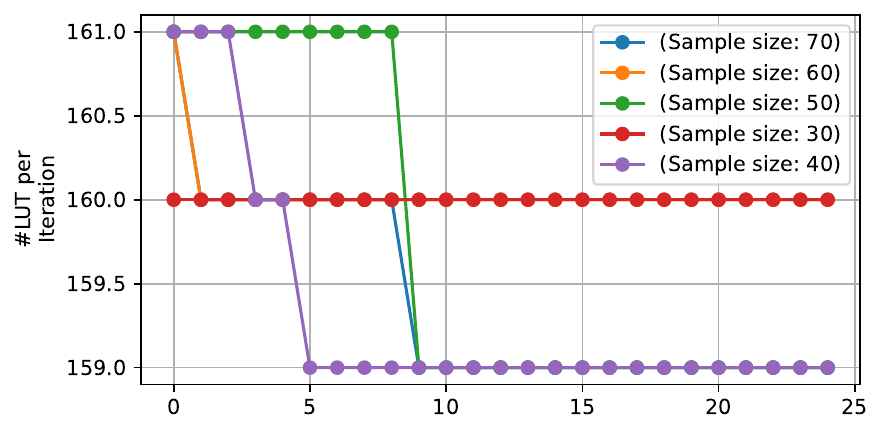}
        \caption{Design: priority}
        \label{fig:priority}
    \end{subfigure}
    \hfill
    \begin{subfigure}[t]{0.24\textwidth}
        \centering
        \includegraphics[width=\textwidth]{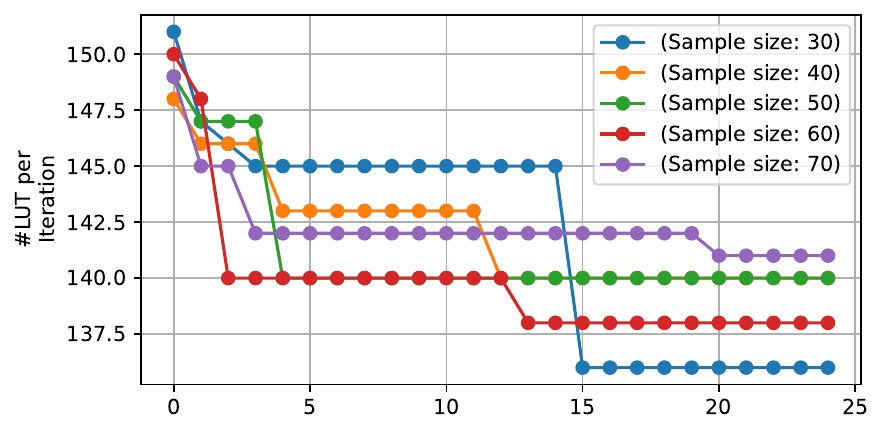}
        \caption{Design: s1488}
        \label{fig:s1488}
    \end{subfigure}
    \hfill
    \begin{subfigure}[t]{0.24\textwidth}
        \centering
        \includegraphics[width=\textwidth]{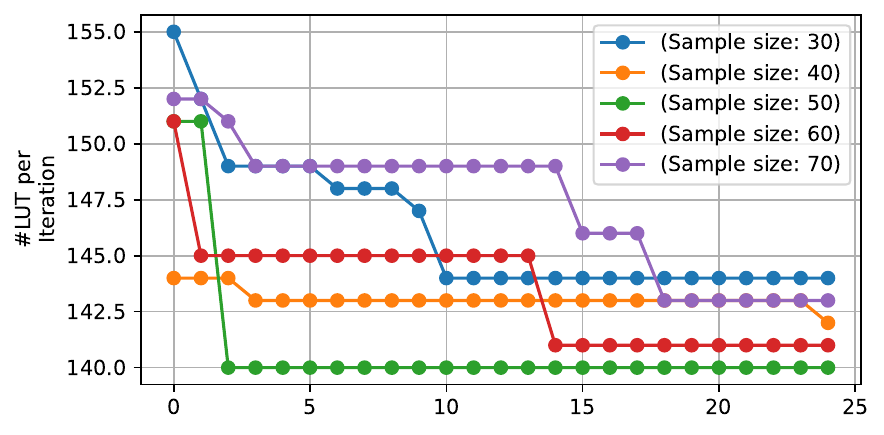}
        \caption{Design: s1494}
        \label{fig:s1494}
    \end{subfigure}
    
    \caption{Convergence analysis of FuseMap with \#LUT size as targeted minimization objective.}
    \vspace{-3mm}
    \label{fig:design_trends}
\end{figure*}

An overview of the FuseMap framework is shown in Figure \ref{fig:flow}. FuseMap is dedicated to addressing a cell selection problem with a pool of $\mathcal{N}$ candidates, denoted as $\mathcal{L}$ (LibSet). Each candidate in $\mathcal{L}$ is associated with two performance metrics: \#LUT and delay. In this work, we use the explored three metrics as reward ($\mathcal{R}$) as a single metric to assess overall circuit efficiency, i.e., \#LUT, delay, and area-delay product (ADP). By definition, $ADP = Delay \times \#LUT$. In each decision iteration, a subset of $n$ candidates is selected from $\mathcal{L}$, forming a potential solution. Each candidate $i \in \mathcal{L}$ corresponds to a binary decision variable $S_i \in \{0,1\}$. A solution is then represented by a multi-hot encoded vector $\mathcal{S} \in \{0,1\}^{\mathcal{N}}$, where $\mathcal{S} = [S_0, S_1, \ldots, S_{\mathcal{N}-1}]$ and $\sum_{i=0}^{\mathcal{N}-1}S_i = n$. Each distinct $\mathcal{S}$ represents a unique combination of candidates, each associated with a specific reward. Here are the essential model formulation configurations:

\textbf{Action Space:} The action space $\mathcal{A}$ consists of a discrete and finite set of actions, where each action corresponds to selecting a specific cell to form a subset of cells from the technology library, subsequently utilized for the technology mapping of a design. The cardinality of this action space is equal to $\mathcal{N}$, denoting the total number of unique cells in the initial technology library. Here, we define each action as $a^i: S_i = 1, i \in [0, \mathcal{N}-1], a^i \in \mathcal{A}$, where taking action $a^i$ means selecting cell $i$ from the original library.

\textbf{State:} During each iteration of forming a subset of cells from the original library with all candidate cells, the state $\mathcal{S}$ represents the current data collection condition, where $\mathcal{S} = [S_0, S_1, \ldots, S_{\mathcal{N}-1}]$. And when $\sum_{i=0}^{\mathcal{N}-1}S_i = n$, where $n$ is the required subset size, it refers to a complete state.

\textbf{Reward function:} The reward function can be flexiblely configured. In this work, we explore three reward functions, i.e., \#LUT, delay, and area-delay-product (ADP) using \#LUT $\times$ delay to explore the area-delay trade-offs. While FuseMap-MAB treats the reward environment as a black-box, it can supports many other complex reward function such as predicted post-routing timing and routing congestion using ML models \cite{yu2019painting, dai2018fast, sohrabizadeh2023robust}.

\textbf{Reward update:} The reward mechanism corresponds to the cell selections made by the agent. We illustrate the reward update using the most complicated case ADP as an example. It is defined as $\mathcal{R}_{\mathcal{S}} = -ADP_{\mathcal{S}} = -\left(\frac{D_{\mathcal{S}}}{D_{Base}} \cdot \frac{A_{\mathcal{S}}}{A_{Base}}\right)$, where $D_{\mathcal{S}}$ and $A_{\mathcal{S}}$ are the metrics derived from the technology mapping of the design using the selected subset of cells indicated at state $\mathcal{S}$. The terms $D_{Base}$ and $A_{Base}$ represent the baseline metrics, established using all the cells in the original library for technology mapping. In this case, $ADP_{\mathcal{S}}$ is a product of normalized $D_{\mathcal{S}}$ and $A_{\mathcal{S}}$. The negative function is employed to invert the metric, i.e., maximizing the reward by minimizing $\mathcal{R}$ to optimize LUT size and/or depth.

Formally, we define the probability vector as $\boldsymbol{p} = [p_{a^0}, p_{a^1}, \ldots, \\p_{a^{\mathcal{N}-1}}]$. \textbf{The probability $p_{a^i}$ is defined as the likelihood of selecting cell $i$ for the sampled library can maximize the reward.} These vectors are updated at each decision epoch based on the observed performance metric, i.e., $\mathcal{R}$. The objective is to iteratively refine $\boldsymbol{p}$ such that the probability of selecting candidates that lead to minimizing $\mathcal{R}$ is maximized.

\subsection{Implementation} We choose to formulate the FuseMap Framework as an Multi-Armed Bandit (FuseMap-MAB) problem. MAB has demonstrate many successes in logic synthesis and technology mapping, with unique advantages of efficiency and performance due to its lightweight model-free reinforcement learning property \cite{yu2020flowtune,neto2022flowtune}. More specifically, our prominent FuseMap-MAB algorithms are with $\epsilon$-greedy strategy \cite{kuleshov2014algorithms} and Upper Confidence Bound strategy \cite{jouini2010upper}, denoted as FuseMap-$\epsilon$ and FuseMap-UCB, respectively. 
As for the bandit problem settings, we refer to each cell $i \in \mathcal{L}$ in the library as an arm $i$, during each iteration, if an action $a^i$ is taken, this means arm $i$ is selected for this action.

\noindent
\textbf{FuseMap-$\epsilon$ Agent.} The FuseMap-$\epsilon$ agent utilizes $\epsilon$-greedy to balance exploration and exploitation via a parameter $\epsilon \in [0, 1]$. This parameter dictates the probability of random action selection (exploration) vs. choosing the action with the highest probability leads to higher reward based on historical data (exploitation). Formally, the agent selects action $a^i \in \mathcal{A}$ according to the following rule:

$a^i = \left\{\begin{matrix}
\argmax_{a^i \in \mathcal{A}} p_{a^i}& \text{with parameter } 1-\epsilon \\ 
\text{a random selection } a^i \in \mathcal{A} & \text{with parameter } \epsilon
\end{matrix}\right.$

\noindent
\textbf{FuseMap-UCB Agent.} The FuseMap-UCB agent integrates a confidence interval around the reward estimates based on historical trial data to tackle the similar exploration-exploitation problem effectively. Action selection is governed by the following formula:
\begin{equation}
    a^i = \argmax_{a^i \in \mathcal{A}} (p_{a^i} + c \sqrt{\frac{\text{log}(t)}{n_{a^i}}})
\end{equation}
where $t$ is the current iteration, $n_{a^i}$ is the number of times that action $a^i$ is taken during $t$ iterations , $c$ is the coefficient that modulates the extent of exploration. Note that, for both FuseMap-$\epsilon$ and FuseMap-UCB agents, the probability vector $\boldsymbol{p}$ are updated as: 
\begin{equation}
    p_{a^i}(t+1) = \frac{p_{a^i}(t) n_{a^i}(t) + \mathcal{R}_{\mathcal{S}}(t)}{n_a(t)}
\end{equation}
where $p_a(t)$ is the probability of action $a^i$ at iteration $t$, $n_{a^i}(t)$ is the number of times that action $a^i$ is taken during $t$ iterations. In this context, the probability of taking action $a^i$ leads to minimizing $\mathcal{R}_{\mathcal{S}}$ is an average of the obtained reward during the data trial process. Note that, we ensure every time when updating $p_{a^i}$ for the next iteration $(t+1)$ at state $\mathcal{S}(t)$, a required number of arms has been selected, so that the reward $\mathcal{R}_{\mathcal{S}}$ is influencing a subset of arms during each iteration.

\section{Results}

 \begin{figure*}[ht]
    \centering
 
        \includegraphics[width=0.98\textwidth]{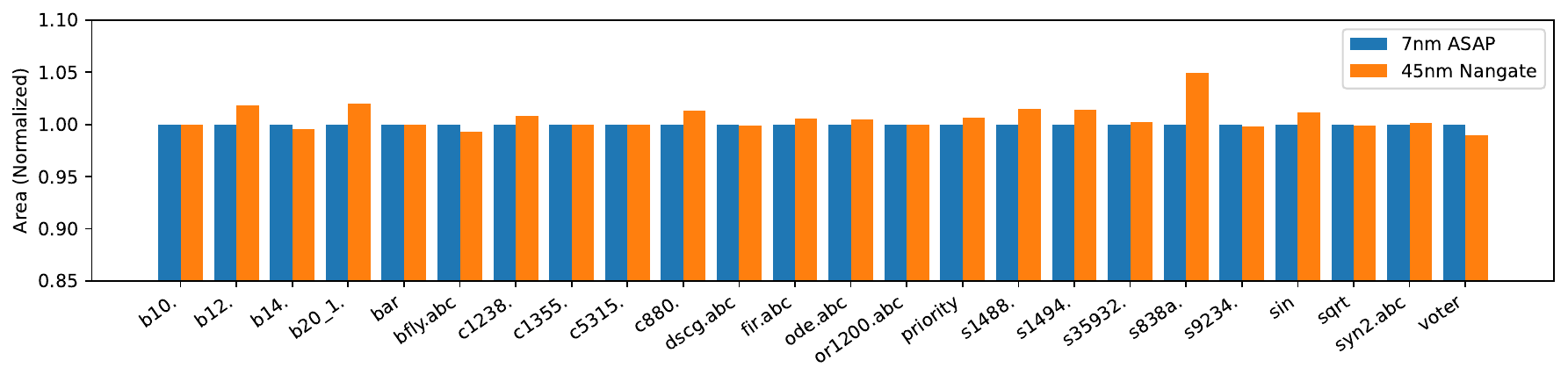}
 \vspace{-2mm}
    \caption{FuseMap performance comparisons between ASAP7 (baseline=1) and NAN45 libraries.}
    \vspace{-3mm}
    \label{fig:library_explore}
\end{figure*}
We evaluate FuseMap from five benchmark suites: ISCAS 85, ISCAS 89, ITC/ISCAS 99, VTR8.0 and EPFL benchmarks, to evaluate the performance of FuseMap after technology mapping and gate sizing are performed. 
All experiments are conducted within ABC framework with an Intel\textregistered Xeon\textregistered Gold 6418H CPU. The vanilla LUT mapping flow invovles standard logic optimization followed by LUT mapping and post-optimization using script \texttt{resyn; strash; dch; strash; resyn; if -K 6; mfs2}. The FuseMap flow involves \texttt{map -a} (area-driven ASIC mapping) right after the \texttt{dch} command, after loading the iteratively updated standard-cell library. Due to the limited space, we exclude the delay-driven mapping FuseMap results as area-driven ASIC mapping performs better.

We evaluate the mapped results with different sampling sizes in FuseMap, which searches for the best achievable results. All experiments are conducted with the 7nm ASAP library \cite{clark2016asap7} and FreePDK45 45nm libary \cite{freepdk45}. For simplicity, we will refer to them as \texttt{ASAP7}, \texttt{NAN45}, respectively. All results presented in this section are conducted with FuseMap framework given a 25 iteration limit with a batch size of 10 in the sampling process. Following the approach used in the motivating case studies, FuseMap is evaluated by fixing its sampling sizes \{30, 40, 50, 60, 70\} over the same technology library among different optimization methods. 

\noindent
\textbf{Runtime of all experiments:} All FuseMap framework performs an iterative tuning guided by MAB, we only report the upper bound runtime. This is because the per iteration runtime varies for different design, due to the size of the design and the variations of mapping process. All experiments finish within 15 mins (\texttt{voter} as the slowest one). For the rest of the section, we structure the results to answer the three following research questions (RQ):


\noindent
\textbf{\underline{RQ1}: How effective is FuseMap in optimizing \#LUT, LUT depth (delay), and ADP?}

\noindent
\textbf{FuseMap demonstrates consistent improvements, particularly effective in \#LUT minimization and ADP.} 
As illustrated in Figure \ref{fig:fuse_map_all}, FuseMap is evaluated across 24 designs using the ASAP7 library, showing noticeable improvements in key metrics: \#LUT, LUT depth (delay), and area-delay product (ADP). The framework’s ability to consistently outperform the baseline in most designs underscores its effectiveness, especially for \#LUT minimization and ADP optimization. In Figure \ref{fig:fuse_map_all}(a), we observe that FuseMap significantly reduces the number of LUTs in 20 out of the 24 designs compared to the baseline. Notable improvements are seen in designs such as \texttt{b12}, where FuseMap reduces the LUT size by approximately 20\% compared to the baseline. Across all designs, the average LUT size improvement is approximately 9\%. 

While the results for LUT depth minimization, shown in Figure \ref{fig:fuse_map_all}(b), are less dramatic than those for LUT size, FuseMap still achieves comparable or better results in most designs. For instance, in the design \texttt{s35932}, FuseMap achieves a reduction in delay of about 32\% compared to the baseline. However, in designs such as \texttt{b20\_1}, FuseMap shows a slight increase in delay, indicating that trade-offs between \#LUT and delay can emerge during the fusion mapping process. On average, FuseMap improves LUT depth (delay) by 3\% across all designs, though a few designs exhibit minimal or no improvement in delay. Finally, across all 24 designs, the average ADP improvement is approximately 9\% (Figure \ref{fig:fuse_map_all}(c)), reflecting FuseMap’s strength in balancing the trade-offs between area and delay during the mapping process. We also apply FuseMap to the entire benchmark suits of ISCAS89, EPFL, and VTR (total 161 designs), which also confirms 9\% ADP improvements on average.

\noindent
\textbf{\underline{RQ2}: Does FuseMap offer stable convergence?}

\noindent
\textbf{FuseMap demonstrates stable convergence trends across all designs and sampling sizes.} 
Illustrated in Figure \ref{fig:design_trends}, the optimization trends for various designs across different sample sizes show that FuseMap consistently converges towards an optimal or near-optimal solution. The convergence behavior varies slightly depending on the design and sample size, but overall, FuseMap exhibits predictable and stable convergence patterns after a few iterations.

The effect of sample size on convergence is notable across the various designs. Smaller sample sizes, such as 30 and 40, often lead to faster convergence and better results in some designs but may vary across different designs. For example, in designs like \texttt{bfly}, \texttt{b12}, and \texttt{ode}, FuseMap converges within the first 5--10 iterations for the smaller sample sizes, as seen in Figures \ref{fig:design_trends}(a)(b)(i). However, these smaller sample sizes can result in a slightly higher \#LUT compared to larger sample sizes, which achieve more refined solutions over time. In designs such as \texttt{b20\_1} and \texttt{s1494}, shown in Figures \ref{fig:design_trends}(d) and \ref{fig:design_trends}(l), larger sample sizes (50 and 70) allow for a more gradual and thorough exploration of the design space, leading to better optimization results.

On average, across all designs and sample sizes, FuseMap converges within approximately 10 to 12 iterations, with faster convergence typically observed for smaller designs and smaller sample sizes. However, the most refined and optimized results are generally achieved with medium to larger sample sizes (40 and 60), as they allow the framework to explore the design space more effectively before settling on a final solution. Importantly, regardless of the sample size, FuseMap does not exhibit any significant divergence or instability during the optimization process, confirming the robustness of the framework across a variety of sampling configurations.


\noindent
\textbf{\underline{RQ3}: Do initial library selections impact FuseMap results?}

\noindent
\textbf{FuseMap performance will vary based on the initial library configuration.} We explore FuseMap using the NAN45 library as the initialized library and compare it with ASAP7 (Figure \ref{fig:library_explore}). While the choice of standard-cell library does not directly affect the design's performance, its combination of cells and area/delay parameters impacts the structure of the mapped gate-level netlist, i.e., one step before LUT mapping in FuseMap. This observation highlights the potential for generating a synthetic standard-cell library that is specifically tuned to improve LUT mapping in fusion mapping scenarios.

\section{Conclusions}

This work presents FuseMap, a novel framework that integrates ASIC technology mapping with LUT mapping to optimize FPGA design results. Our approach leverages the strengths of modern ASIC technology mapping flows and introduces a lightweight reinforcement learning-based mechanism to fine-tune technology libraries for improved LUT mapping performance. Through extensive case studies and evaluations, we demonstrated the potential of FuseMap to consistently outperform traditional LUT mapping techniques. 
FuseMap achieved an average improvement of 8\% in LUT minimization and a 9\% improvement in ADP, demonstrating its effectiveness in balancing the trade-offs between area and delay. While the improvements in delay were more modest, with an average reduction of 3\%, FuseMap generally maintained or improved delay performance across most designs.

\textbf{Acknowledgment} This work is sponsored by the National Science Foundation under Grant No. CCF2403134, CCF2349670, CNS2349461, CNS2229562, and DARPA under Grant No. HR001125C0058.




\bibliographystyle{unsrt}
\bibliography{tex/synthesis}





\end{document}